\documentclass[a4paper,10pt,abstracton,openbib,final]{scrartcl}
\usepackage{color}
\usepackage[pdftex]{graphicx}
\usepackage{amsmath}
\usepackage{amsfonts}
\usepackage{amssymb, amsxtra}
\usepackage{mathrsfs}
\usepackage{float}
\usepackage{bm, upgreek}
\usepackage{subfigure}
\usepackage{booktabs, multirow, cite}
\usepackage{authblk, caption}

\DeclareMathAlphabet{\mathscr}{OT1}{pzc}{m}{it}
\newcommand{\begeq}{\begin{equation}\begin{gathered}}
\newcommand{\eqend}{\end{gathered}\end{equation}}
\newcommand{\begal}{\begin{equation}\begin{aligned}}
\newcommand{\alend}{\end{aligned}\end{equation}}

\renewcommand{\t}[1]{\bm{#1}}

\renewcommand{\d}{\, \mathrm d }

\newcommand{\p}{\partial}

\newcommand{\del}{\updelta}

\newcommand{\pd}[2]{\frac{\p #1}{\p #2}}

\newcommand{\comma}{\ , \ \ }

\newcommand{\cen}{\overset{\text{c}}}
\title{\huge Determination of Metamaterial Parameters by Means of a Homogenization Approach Based on Asymptotic Analysis}
\author{
H. Yang\thanks{Technische Universit\"at Berlin, Institute of Mechanics, Berlin, Germany}
,
B. E. Abali\thanks{Technische Universit\"at Berlin, Institute of Mechanics, Berlin, Germany}
,
W. H. M\"uller\thanks{Technische Universit\"at Berlin, Institute of Mechanics, Berlin, Germany}
, and
D. Timofeev 
\thanks{Universit\'a degli Studi dell'Aquila, Dipartimento di Ingegneria e Scienze dell'Informazione e Matematica, L'Aquila, Italy}	
}
\date{}

\begin{document}
\maketitle

\begin{abstract}
Owing to additive manufacturing techniques, a structure at millimeter length scale (macroscale) can be produced by using a lattice substructure at micrometer length scale (microscale). Such a system is called a metamaterial at the macroscale as the mechanical characteristics deviate from the characteristics at the microscale. As a remedy, metamaterial is modeled by using additional parameters; we intend to determine them. A homogenization approach based on the asymptotic analysis establishes a connection between these different characteristics at micro- and macroscales. A linear elastic first order theory at the microscale is related to a linear elastic second order theory at the macroscale. Relation for parameters at the macroscale is derived by using the equivalence of energy at macro- and microscales within a so-called  Representative Volume Element (RVE). Determination of parameters are succeeded by solving a boundary value problem with the Finite Element Method (FEM). The proposed approach guarantees that the additional parameters vanish if the material is purely homogeneous, in other words, it is fully compatible with conventional homogenization schemes based on spatial averaging techniques. Moreover, the proposed approach is reliable as it ensures that such resolved additional parameters are not sensitive to choices of RVE consisting in the repetition of smaller RVEs but depend upon the intrinsic size of the structure. 
\end{abstract}

\section{Introduction}
Periodic lattice type structures involving large number of repetitive substructures continue to attract the interest of many researchers because of their fascinating properties like relatively low manufacturing costs, high specific stiffness, etc. \cite{gibson2005biomechanics, gibson1999cellular, barchiesi2018mechanical, lu2017modelling,giorgio2017influence,mroz1981optimal}.  Mechanical respons of such a structure depends not only on the material, but also on the morphology of its substructure \cite{liu2017topology, liu2019manufacturing}. Hence, metamaterial is used for such a substructure depending material. 

In order to design and fabricate metamaterials for engineering applications, the accurate and efficient prediction of their mechanical performances is of importance \cite{tekouglu2008size,chen2002size,dell2011variational,turco2016large,hendy2008numerical}. Indeed, standard numerical  techniques like the Finite Element Method (FEM) are utilized to achieve a modeling of such a structure with every detail of its substructure \cite{yang2018material,Yang2019Computation}. However, it requires the mesh size to be at least one order smaller than the substructure geometric size leading to very high computational costs. Hence, homogenization techniques are developed to upscale the mechanical response at the microscale---the presence of the substructure leads to a composite material, which can be seen as a heterogeneous material---to the macroscale by defining an appropriate constitutive equation. Especially in composite materials, with fibers embedded in a matrix building a periodic substructure, micro- and macroscale behaviors are modeled by the same linear elastic model, also called a \textsc{Cauchy} continuum.  The homogenization of such periodic structures toward an equivalent \textsc{Cauchy} continuum has been investigated thoroughly \cite{ghosh1996two,kushnevsky1998identification,bensoussan2011asymptotic,noor1988continuum,zohdi2017homogenization,nazarenko2018effective}. 

Many approaches in the literature assume that there exists an Representative Volume Element (RVE) with periodic boundary conditions that precisely captures the deformation behavior of the whole geometry. Such an approach utilizes the energy equivalence of the RVE at both macroscale and microscale as also used in \cite{hill1972constitutive}.  The effective properties of such homogenized continua is in good agreement with experiments \cite{sun1996prediction} under the condition that $L\gg l$, where $L$ represents the macroscopic length scale, i.e., (mean value of ) the geometric dimensions of the whole structure, and $l$ represents the length scale of the microscale, namely, the geometric dimensions of the substructure. The quantity $l$ will be used as the ``length scale'' of a basic cell of the structure, as demonstrated in Fig.\,\ref{Yang_Homo_1}. Note that the concept of a basic cell is different from an RVE. It is evident that a basic cell can be regarded as an RVE, and stacking or gathering several basic cells can construct an RVE as well. Classical homogenization encounters limitations  \cite{li2011micromechanics,askes2011gradient} when $L$ is of a comparable order of scale with respect to $l$. 

Size effects fail to be captured by a standard homogenization having the same order theory at both scales. A feasible approach is to use first order at the microscale and second order at the macroscale leading to additional parameters at the macroscale to be determined, we refer to various formulations of a second order theory in \cite{mindlin1968first, toupin1962elastic, altenbach2009linear, 030, 021, dell2009generalized, eremeyev2012material, rosi2013propagation, placidi2018two, placidi2018strain, spagnuolo2017qualitative, andreaus2018ritz, placidi2017inverse, placidi2018simulation, rosi2018validity, altenbach2008direct, abali2018revealing}. Higher order theories are called \textit{generalized mechanics} and homogenization in the framework of generalized mechanics is a challenging task endeavored by many scientists, among others by \cite{forest1999estimating, pietraszkiewicz2009natural, giorgio2017continuum, dell2016large, rahali2015homogenization, kouznetsova2002multi, barchiesi2018out}. Mostly, it is agreed that homogenization of an RVE by involving so-called higher gradient terms of the macroscopic field is a natural way to include size effect \cite{bacigalupo2018identification,li2011micromechanics, forest2001asymptotic,franciosi2018mean, franciosi2004using,franciosi2011effective}. Using gamma-convergence, homogenization results have been obtained in \cite{alibert2003truss,pideri1997second,alibert2015second}. A remarkable class of structures being described at the macroscale by using a second gradient elasticity theory in pantographic structures \cite{barchiesi2017review, steigmann2015mechanical, scerrato2016three}, which have received a notable follow-up in the literature \cite{de2019macroscopic, dell2018pantographic, placidi2016review, turco2017pantographic, turco2016fiber}, also from a mathematically rigorous standpoint, regarding fundamental issues such as well-posedness \cite{eremeyev2017linear}.

A possibly promising homogenization technique is the asymptotic analysis, which has been used to obtain homogenized material parameters in \cite{smyshlyaev2000rigorous}. This method decomposes variables to their global variations and local fluctuations. Such a decomposition is used to generate closed form equations to determine constitutive parameters as applied in one-dimensional problems, for example in the analysis of composites \cite{boutin1996microstructural, barchiesi20181d}, while 2D problems \cite{bacigalupo2014second, barboura2018establishment, boutin2017linear, placidi2015gedanken, cuomo2014variational} have been investigated numerically.  FEM is employed in \cite{peerlings2004computational} demonstrating that higher-order terms start dominating as the difference between parameters of composite materials increases. A second-order asymptotic and computational homogenization technique is proposed by \cite{bacigalupo2014second} that solves the boundary value problems generated by the asymptotic homogenization with a quadratic ansatz. However, there are still two main issues which are not well discussed when trying to homogenize structures in the framework of generalized mechanics \cite{tran2012micromechanics}:
\begin{itemize}
\item
The first one concerns the compatibility such that parameters of the strain gradient stiffness tensor should vanish when the structure is purely homogeneous. 
\item
The second one is about reliability such that the strain gradient stiffness tensor has to be insensitive to the repetition of the basic cell.
\end{itemize}
A successful attempt is made in \cite{li2011establishment,li2011micromechanics} establishing a connection between microscale parameters and macroscale parameters (by using the strain gradient theory) by proposing a ``correction'' term rendering the strain gradient stiffness tensor satisfying compatibility and reliability requirements. Different numerical solution methods are used for this approach, Fast Fourier Technique (FFT) is employed in \cite{li2013numerical} and FEM is exploited in \cite{barboura2018establishment}. We follow their methodology and propose an alternative derivation for this ``correction'' term in Section~\ref{sec:2} and try to do it in a pedagogical way. Furthermore, we apply and validate the method for simple yet general 2D metamaterials in Sections~\ref{sec:3} and \ref{sec:4} by using the FEM. In order to demonstrate its versatility, computations of the square lattice are performed in Section~\ref{sec:5}. The computations are performed with the aid of open-source codes developed by the FEniCS project \cite{030}. The proposed method delivers all metamaterial parameters in 2D by using linear elastic material model at the microscale after a computational procedure as investigated in the following.

\section{Connection of micro- and macroscale parameters} \label{sec:2}

\begin{figure}
	\centering
	\includegraphics[width=0.55\textwidth]{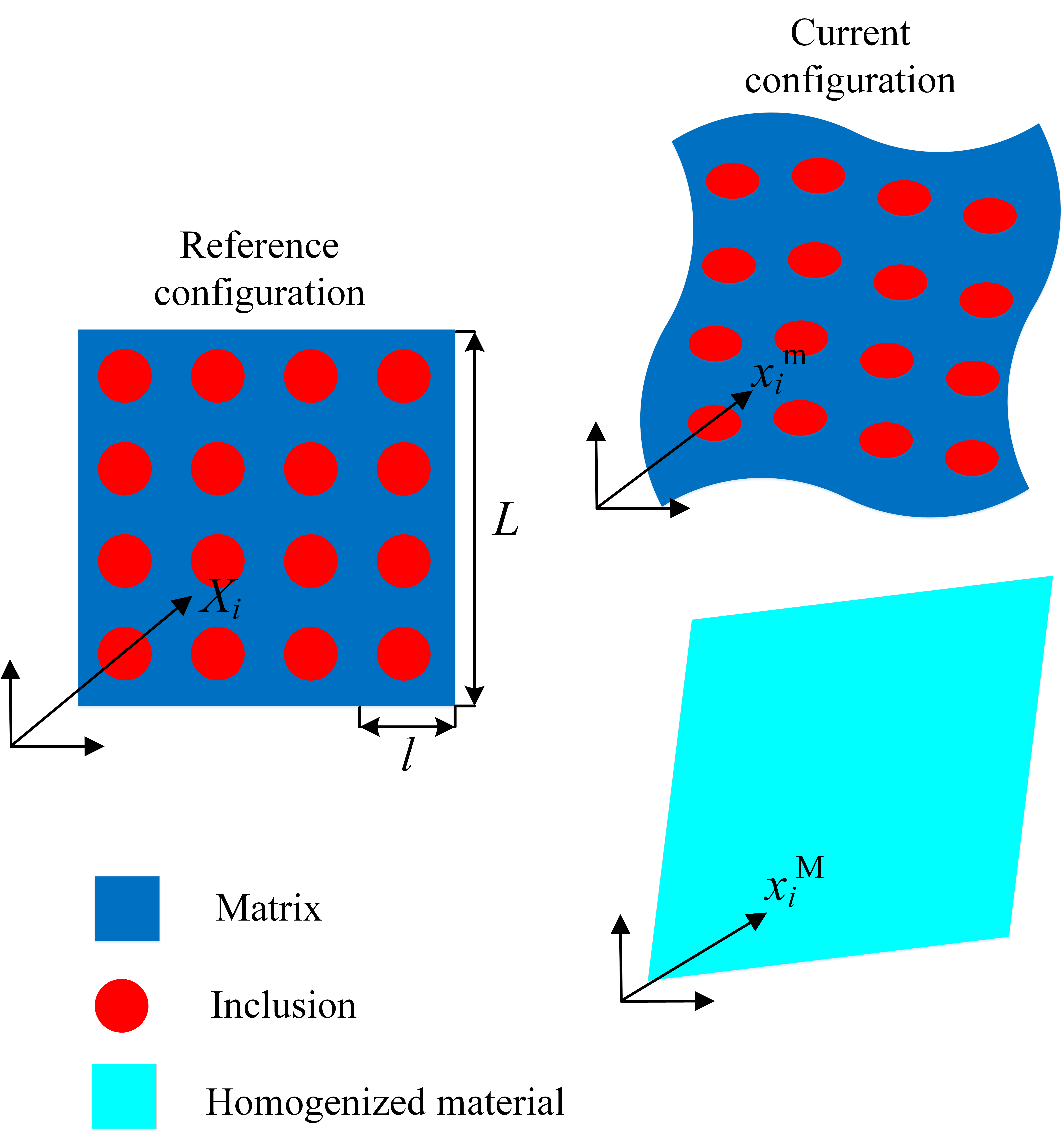}
	\caption{Left: continuum body in the reference frame. Right top: deformation at the microscale. Right bottom: corresponding deformation at the macroscale.}
	\label{Yang_Homo_1} 
\end{figure}

Consider a continuum body occupying a domain $\Omega$ in two-dimensional space, $\Omega \in \mathbb R^2$. The metamaterial embodies an RVE, $\Omega^P$, where periodically aligned RVEs constitute metamaterials domain,
\begin{equation} 
\cup \Omega^P = \Omega   \comma 
\Omega^P \cap \Omega^Q = \emptyset  \comma
 \qquad P,Q = 1,2,3,...M, P \neq Q \ .
\end{equation}
RVE at the microscale represents the detailed substructure like fibers and matrix in a composite material. The same RVE at the macroscale is modeled by a homogeneous metamaterial and we assume that their stored energy values are equivalent although the definitions at both scales differ. We use a first order theory for defining the energy (volume) density of an RVE at the microscale, $w^\text{m}$, whereas we utilize a second order theory at the macroscale for the energy density, $ w^\text{M}$, leading to
\begeq  \label{equivalence of energy}
\int_{\Omega^P} { w^\text{m}} \d V = \int_{\Omega^P} { w^\text{M}} \d V \ ,\\
\int_{\Omega^P} \frac{1}{2} C^\text{m}_{ijkl}  u^\text{m}_{i,j}  u^\text{m}_{k,l} \d V = \int_{\Omega^P} \frac{1}{2} \big(  C^\text{M}_{ijkl}  u^\text{M}_{i,j}  u_{k,l}^\text{M} +   D_{ijklmn}^\text{M}  u_{i,jk}^\text{M}  u_{l,mn}^\text{M} \big) \d V \ ,
\eqend
where and henceforth we apply \textsc{Einstein} summation convention over repeated indices and use a comma notation for space derivatives in $\t X$. Moreover, all fields are expressed in Cartesian coordinates. The microscale stiffness tensor, $C^\text{m}_{ijkl}$, is a function in space. Consider a lattice substructure. Even if the trusses are of homogeneous material, voids between trusses generates a heterogeneous substructure at the microscale such that microscale stiffness tensor depends on space coordinates and possesses either the value of truss material or zero due to voids. In contrary, macrocale material tensors, $C^\text{M}_{ijkl}$ and $D^\text{M}_{ijklmn}$, are constant in space as they are generated by the homogenization procedure to be explained in the following. The continuum body at the reference frame has particles at coordinates $X_i$, where they move to $x_i$ under a mechanical loading. The displacement is the deviation from the reference frame and we emphasize that the microscale displacement, $u^\text{m}_i$, is different than the macroscale displacement, $u^\text{M}_i$,
\begeq \label{AY_eq05}
{u^\text{m}_i} = { x_i^\text{m}} - X_i \ , \\
{u^\text{M}_i} = { x_i^\text{M}} - X_i \ ,
\eqend
as the current positions of particles differ. This difference between ${x_i^\text{m}}$ and ${ x_i^\text{M}}$ is illustrated in Fig.\,\ref{Yang_Homo_1}. For demonstrating the microscale deformation, the substructure is visualized as well. For simplicity, a well-known example is used, namely, composite materials with the red inclusion (fibers) embedded in the blue material (matrix). For the homogenized case, the same particle moves to $ x_i^\text{M}$ expressed at the macro-scale without the substructure. We emphasize that micro- and macroscales are both expressed in the same coordinate system. Two different cases are examined, a heterogeneous case at the microscale with known material properties versus a homogeneous case at the macroscale with sought parameters. In order to identify the material parameters, the strain energy expressions for macro- and microscales are derived in what follows.

\subsection{The macroscale energy for an RVE}

Consider the macroscale case for an RVE, $\Omega^P$. As usual in spatial averaging, we define the geometric center $\cen{\t X}$ of the RVE,
\begeq \label{center_def}
\cen{\t X}  = \frac1V \int_{\Omega} \t{X} \d V \ ,
\eqend
and approximate the macroscale displacement by a \textsc{Taylor} expansion around the value at the geometric center by truncating after quadratic terms (in order to account for the strain gradient effect) and calculate displacement gradients of this approximation
\begal  \label{Taylor's series of displacements}
{u^\text{M} _i}(\t{X}) &= {u^\text{M} _i}\Big|_{\cen{\t X}} 
+  u^\text{M}_{i,j} \Big|_{\cen{\t X}} (X_j-\cen X_j) 
+ \frac{1}{2}  u^\text{M}_{i,jk} \Big|_{\cen{\t X}} (X_j-\cen X_j) (X_k-\cen X_k)  
\ , \\
u^\text{M}_{i,l}(\t{X}) &= u^\text{M}_{i,j} \Big|_{\cen{\t X}} \delta_{jl}
+ \frac{1}{2}  u^\text{M}_{i,jk} \Big|_{\cen{\t X}} ( \delta_{jl} (X_k-\cen X_k) + (X_j-\cen X_j) \delta_{kl} ) 
\ , \\
&= u^\text{M}_{i,l} \Big|_{\cen{\t X}} +  u^\text{M}_{i,lk} \Big|_{\cen{\t X}} (X_k-\cen X_k)
\ , \\
u^\text{M}_{i,lm}(\t{X}) &= u^\text{M}_{i,lk} \Big|_{\cen{\t X}} \delta_{km} = u^\text{M}_{i,lm} \Big|_{\cen{\t X}} \ .
\alend
According to Eq.~\eqref{Taylor's series of displacements}, spatial averaging the gradient terms of the displacement field reads
\begal \label{displacement average}
	\langle {u}_{i,j}^\text{M} \rangle &=\frac{1}{V}\int_{\Omega^P} u^\text{M} _{i,j} \d V= u^\text{M}_{i,j}\Big|_{\cen{\t X}} + u^\text{M}_{i,jk}\Big|_{\cen{\t X}} \bar I_k 
	\comma
	\bar I_k = \frac1V \int_{\Omega^P} (X_k-\overset{\text{c}}X_k) \d V
	\ , \\
	\langle {u}_{i,jk}^\text{M} \rangle &=\frac{1}{V}\int_{\Omega^P} u^\text{M} _{i,jk}\d V= u^\text{M}_{i,jk} \Big|_{\cen{\t X}} \ .
\alend
As we acquire $\bar I_k = 0 $ from Eq.\,\eqref{center_def},
\begeq \label{displacement average approx}
	\langle {u}_{i,j}^\text{M} \rangle =  u^\text{M}_{i,j}\Big|_{\cen{\t X}} \comma
	\langle {u}_{i,jk}^\text{M} \rangle = u^\text{M}_{i,jk} \Big|_{\cen{\t X}} \ .
\eqend
After inserting Eq.~\eqref{displacement average approx} into Eq.~\eqref{Taylor's series of displacements}, we obtain
\begal \label{dis and grad dis}
	u_i^\text{M}(\t X) &= u^\text{M} _i\Big|_{\cen{\t X}} + \langle {u}_{i,j}^\text{M} \rangle (X_j-\overset{\text{c}}X_j) +  \frac{1}{2} \langle {u}_{i,jk}^\text{M} \rangle (X_j-\overset{\text{c}}X_j) (X_k-\overset{\text{c}}X_k)  \ , \\
	u^\text{M} _{i,j}(\t X) &= \langle{u}_{i,j}^\text{M} \rangle + \langle{u}_{i,jk}^\text{M}\rangle (X_k-\overset{\text{c}}X_k) \ , \\
	u^\text{M}_{i,jk}(\t X) &= \langle {u}_{i,jk}^\text{M}  \rangle \ .
\alend
Now, by using the latter Eq.~\eqref{dis and grad dis} on the right-hand side of Eq.~\eqref{equivalence of energy}, macroscale energy of an RVE is acquired as macroscale tensors are constant in space
\begeq \label{strain energy for homogenized material} 
\int_{\Omega^P} \frac{1}{2} \big(  C^\text{M}_{ijlm} u^\text{M}_{i,j} u^\text{M}_{l,m} +  D^\text{M}_{ijklmn} u^\text{M}_{i,jk} u^\text{M}_{l,mn} \big) \d V 
= \frac12 C^\text{M}_{ijlm}  \int_{\Omega^P} u^\text{M}_{i,j} u^\text{M}_{l,m} \d V
+ \\
\frac12 D^\text{M}_{ijklmn} \int_{\Omega^P} u^\text{M}_{i,jk} u^\text{M}_{l,mn} \d V 
= \frac{1}{2} C^\text{M}_{ijlm} \int_{\Omega^P} \Big( \langle{u}_{i,j}^\text{M} \rangle +\langle {u}_{i,jk}^\text{M} \rangle (X_k-\cen X_k) \Big) 
\times \\
\times \Big(\langle{u}_{l,m}^\text{M} \rangle + \langle{u}_{l,mn}^\text{M}\rangle (X_n-\cen X_n) \Big) \d V +   \frac12 D^\text{M}_{ijklmn} \int_{\Omega^P} \langle{u}_{i,jk}^\text{M}\rangle \langle{u}_{l,mn}^\text{M}\rangle  \d V 
= \\
=  \frac{1}{2}V \Big(  C^\text{M}_{ijlm} \langle{u}_{i,j}^\text{M}\rangle \langle{u}_{l,m}^\text{M}\rangle +  ( C^\text{M}_{ijlm} \bar{I}_{kn} +  D^\text{M}_{ijklmn} ) \langle{u}_{i,jk}^\text{M}\rangle \langle{u}_{l,mn}^\text{M}\rangle \Big) \ , \\
\eqend
where 
\begeq
\bar{I}_{kn} = \frac1V \int_{\Omega^P} (X_k-\cen X_k)(X_n-\cen X_n) \d V \ .
\eqend
Consequently, the macroscale energy of an RVE is expressed in terms of the gradient of macroscopic deformation. In what follows, it will be shown, by employing asymptotic homogenization analysis, that the microscale energy can be formulated in terms of the gradient of macroscopic deformation as well leading to connections between parameters.

\subsection{The microscale energy for an RVE}

By following the asymptotic homogenization method as in \cite{pinho2009asymptotic}, we reformulate the left-hand side of Eq.~\eqref{equivalence of energy}. The asymptotic homogenization method separates length scales by using global coordinates, $\t X$, for describing the global variation of the displacement, and by using local coordinates, $\t y$, for describing the local fluctuation of the displacement. We refer to \cite{peszynska2007multiscale} and \cite[Appendix B]{efendiev2009multiscale} for a more detailed investigation of the multiscale asymptotic analysis applied herein. We introduce the local coordinates,
\begin{equation} \label{link_coord}
y_j= \frac1\epsilon ( X_j - \cen X_j ) \ ,
\end{equation}
where $\epsilon$ is a homothetic ratio scaling global and local coordinates. We stress that the dimensions of an RVE in local coordinates can be arbitrarily chosen by varying $\epsilon$. 
\begin{figure}
	\centering
	\includegraphics[width=0.8\textwidth]{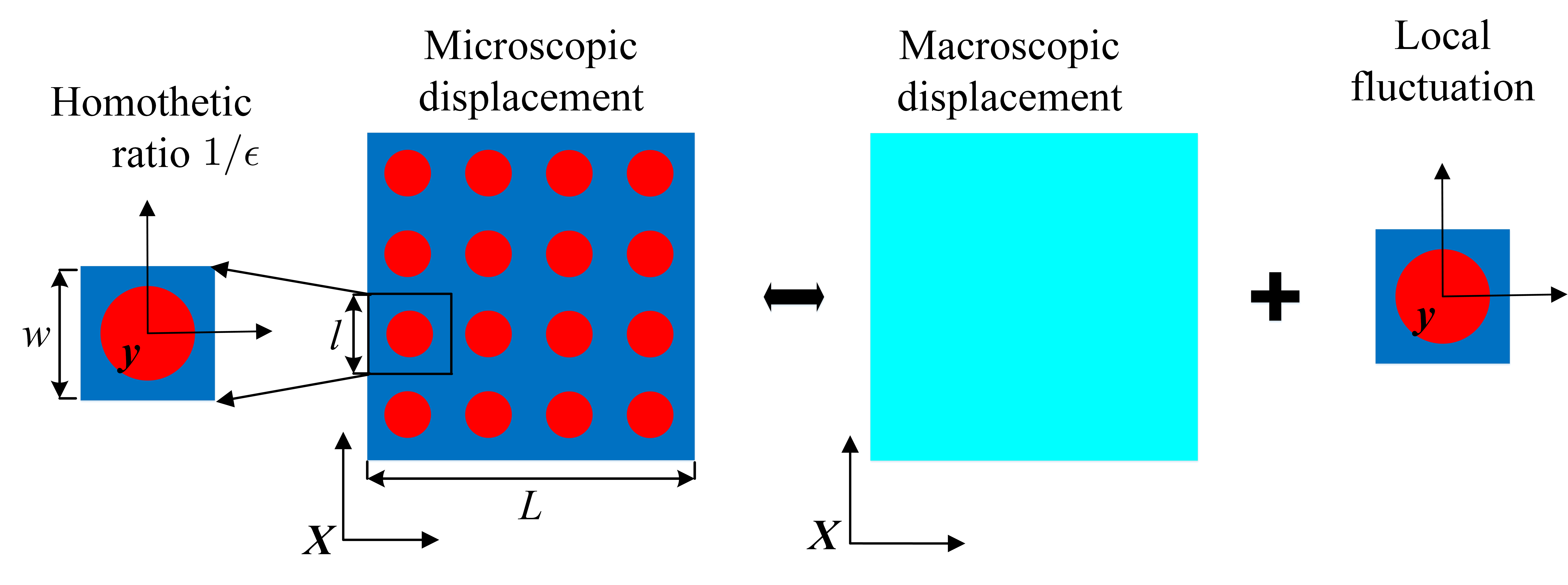}
	\caption{Illustration of the approximation of the asymptotic expansion.}
	\label{Yang_Homo_2} 
\end{figure}
For example, as depicted in Fig.\,\ref {Yang_Homo_2}, the size of an RVE is given by $l$ in global coordinates, whereas it is denoted by $w$ in local coordinates. If we choose $l=0.001$\,mm, as measured in global coordinates, $\t X$, then it can be homothetically scaled to any dimension, such as $w=0.001$\,mm or $w=1000$\,mm in local coordinates, $\t y$, by setting the homothetic ratio to $ \epsilon = 1.0 $ or $\epsilon = 10^{-6}$, in such a way that the size of the RVE is kept constant in the global coordinates. We remark that the homothetic ratio is used to describe the relationship for the sizes of an RVE between global and local coordinates; but the ratio between macroscale and microscale remains the same, $L/l=\text{const}$. We assume that the displacement field is a smooth function at the macroscopic level and $\t y$-periodic in local coordinates resulting in vanishing mean local fluctuations within each RVE. Hence, the decomposition of the microscale displacement is additive into macroscale displacement and local fluctuations defined in different scales---they are independent.

By following \cite{li2013numerical}, the displacement field of an RVE, $\Omega^P$, at global coordinates $\t{X}$ is expanded by using asymptotic series with homothetic ratio $\epsilon$, where the corresponding coefficients in general do depend on global coordinates, $\t{X}$, as well as local coordinates, $\t{y}$, which are related by Eq. (\ref{link_coord}),
\begin{equation} \label{displacement function}
\t{u ^\text{m}}(\t X) = \overset{0} {\t u}( \t X, \t y) + \epsilon \overset{1}{\t u}(\t X,\t y) + \epsilon^2 \overset{2} {\t u}(\t X, \t y) + \dots \ ,
\end{equation}
where $\overset{n} {\t u}( \t X, \t y)$ ($n$ = 0, 1, 2, \dots) are assumed to be $\t y$-periodic. We shall see later that the first term $\overset{0} {\t u}( \t X, \t y)$ is independent of $\t y$. We apply now the elasticity problem in statics as it  needs to be fulfilled within the RVE 
\begeq \label{local govening equation}
\big( C_{ijkl}^\text{m}  u^\text{m}_{k,l} \big)_{,j} + f_{i} =  0 \qquad \forall \t X \in \Omega^P \ ,
\eqend
where the body force, $\t f$, is a given function. By inserting Eq.\,\eqref{displacement function} as well as using the chain rule with the aid of the relation in Eq.\,\eqref{link_coord}, we obtain
\begeq
u^\text{m}_{i,j} 
= \Big( \overset{0} u_i( \t X, \t y) + \epsilon \overset{1}u_i (\t X,\t y) + \epsilon^2 \overset{2}u_i(\t X, \t y) + \dots \Big)_{,j}
= \\
= \overset{0} u_{i,j} + \pd{\overset{0} u_i}{y_k} \frac{\delta_{kj}}{\epsilon} 
+ \epsilon \overset{1}u_{i,j} + \epsilon \pd{\overset{1}u_i}{y_k} \frac{\delta_{kj}}{\epsilon}  
+ \epsilon^2 \overset{2}u_{i,j} + \epsilon^2 \pd{\overset{2}u_i}{y_k} \frac{\delta_{kj}}{\epsilon}  + \dots
\eqend
Using the latter in Eq.\,\eqref{local govening equation} 
\begeq
 \bigg( C_{ijkl}^\text{m}  \Big( 
\overset{0} u_{k,l} 
+ {\frac1{\epsilon}} \pd{\overset{0} u_k}{y_l} 
+ \epsilon \overset{1}u_{k,l} 
+  \pd{\overset{1}u_k}{y_l} 
+ \epsilon^2 \overset{2}u_{k,l} 
+ \epsilon \pd{\overset{2}u_k}{y_l} 
 \Big) \bigg)_{,j}
+ \\
+ \pd{}{y_j}\bigg( C_{ijkl}^\text{m}  \Big( 
{\frac1{\epsilon}} \overset{0} u_{k,l} 
+ {\frac1{\epsilon^2}} \pd{\overset{0} u_k}{y_l} 
+  \overset{1}u_{k,l} 
+ {\frac1{\epsilon}} \pd{\overset{1}u_k}{y_l} 
+ \epsilon \overset{2}u_{k,l} 
+   \pd{\overset{2}u_k}{y_l} 
 \Big) \bigg) 
 + f_{i} =  0 
\eqend
and then gathering terms having the same order in $\epsilon$ leads to the following terms:
\begin{itemize}
\item
in the order of $\epsilon^{-2}$
\begin{equation} \label{first terms}
\frac{\partial }{\partial  y_j} \Big( C_{ijkl}^\text{m}   \frac{\partial \overset{0}u_k}{\partial y_l} \Big) = 0 \ ;
\end{equation}
\item
in the order of $\epsilon^{-1}$
\begin{equation} \label{second terms}
\Big( C_{ijkl}^\text{m} \frac{\partial \overset{0}u_k}{\partial y_l} \Big)_{,j} 
+  \frac{\partial }{\partial  y_j} \big( C_{ijkl}^\text{m}   \overset{0}u_{k,l} \big)  
+ \frac{\partial }{\partial  y_j} \Big( C_{ijkl}^\text{m}   \frac{\partial \overset{1}u_k}{\partial y_l} \Big)   
= 0 \ ;
\end{equation}
\item
in the order of $\epsilon^{0}$
\begeq \label{third terms}
\big( C_{ijkl}^\text{m}   \overset{0}u_{k,l} \big)_{,j} 
+  \Big( C_{ijkl}^\text{m}   \frac{\partial \overset{1}u_k}{\partial y_l} \Big)_{,j}
+  \frac{\partial }{\partial  y_j} \big( C_{ijkl}^\text{m}  \overset{1}u_{k,l} \big)
+ \frac{\partial }{\partial  y_j} \Big( C_{ijkl}^\text{m}  \pd{\overset{2}u_k}{y_l} \Big)   
+ f_i = 0 \ .
\eqend
\end{itemize}
By solving these partial differential equations, Eq.~\eqref{displacement function} can be rewritten as
\begin{equation} \label{displacement function 0}
u^\text{m}_i(\t X,\t y) 
= \overset{0}{u}_i(\t X) 
+ \epsilon \varphi_{abi}(\t y) \overset{0}{u}_{a,b}(\t X) 
+ \epsilon^2 \psi_{abci}(\t y) \overset{0}{u}_{a,bc}(\t X) 
+ \dots \ .
\end{equation}
in which $\varphi_{abi}(\t y)$ and $\psi_{abci}(\t y)$ are both $\t y$-periodic and they are the solutions of the following two partial differential equations: 

\begeq  \label{diff1}
\frac{\partial }{\partial  y_j} \bigg( C_{ijkl}^\text{m} \Big(  \frac{\p \varphi_{abk}}{\p y_l} + \delta_{ak} \delta_{bl} \Big) \bigg)
= 0 \ ,
\eqend
\begeq \label{diff2}
\frac{\partial }{\partial  y_j} \bigg( C_{ijkl}^\text{m} \Big(  \frac{\p \psi_{abck}}{\p y_l} +  \varphi_{abk} \delta_{lc} \Big) \bigg)
+  C_{ickl}^\text{m} \Big(  \frac{\p \varphi_{abk}}{\p y_l} + \delta_{ka} \delta_{lb}  \Big) - {C}_{icab}^\text{M} = 0 \ .
\eqend
A note should be made that the choice of the indices of the third order tensor $\t \varphi$ and fourth order tensor $\t \psi$ differs from those in \cite{li2013numerical, barboura2018establishment}. Since $\t \varphi$ and $\t \psi$ are expressed in the Cartesian coordinates, we choose to use lower indices like  $\varphi_{abk}$ and $ \psi_{abck}$ in the context, and they are mathematically and physically exactly identical to those in \cite{li2013numerical, barboura2018establishment}.
We refer to Appendix for a derivation of Eq.~\eqref{displacement function 0}, Eq.~\eqref{diff1} and Eq.~\eqref{diff2}.
Since the first term $\overset{0}{u}_i(\t X)$ depends only on the macroscopic coordinates, $\t{X}$, it is assumed to be the known macroscopic displacement $\overset{0}{u}_i(\t X) =u^\text{M}_i(\t X)$ such that Eq.~\eqref{displacement function} provides
\begin{equation} \label{displacement function macro}
u^\text{m}_i(\t X,\t y) 
= {u}_i(\t X)^\text{M} 
+ \epsilon \varphi_{abi}(\t y) {u}_{a,b}^\text{M}(\t X) 
+ \epsilon^2 \psi_{abci}(\t y) {u}_{a,bc}^\text{M}(\t X) 
+ \dots \ .
\end{equation}

We aim at defining the energy at microscale, thus, we need the gradient of the microscale displacement,
\begal 
u^\text{m}_{i,j} &= 
\Big( u_i^\text{M} + 
	\epsilon \varphi_{abi} u_{a,b}^\text{M} + 
	\epsilon^2 \psi_{abci} u_{a,bc}^\text{M} + \dots \Big)_{,j}  \\
	&= u_{i,j}^\text{M} 
	+ \frac{\p \varphi_{abi}}{\p y_j}  u^\text{M}_{a,b}
	+ \epsilon \varphi_{abi} u_{a,bj}^\text{M}
	+ \epsilon \frac{\p \psi_{abci}}{\p y_j} u^\text{M}_{a,bc} + \epsilon^2 \psi_{abci} u^\text{M}_{a,jbc} + \dots 
\alend
with the same accuracy, i.e., after neglecting higher than second gradients and inserting Eq.\,\eqref{dis and grad dis} with the aid of Eq.\,\eqref{link_coord}
\begal \label{nabla u}
u^\text{m}_{i,j}
	&= \Big(\delta_{ia} \delta_{jb} + \frac{\p \varphi_{abi}}{\p y_j} \Big) u^\text{M}_{a,b} 
	+ \epsilon u^\text{M}_{a,bc} \Big( \varphi_{abi} \delta_{jc} + \frac{\p \psi_{abci} }{\p y_j} \Big) + \dots \\
	&= \Big(\delta_{ia} \delta_{jb} + \frac{\p \varphi_{abi}}{\p y_j} \Big)  
	\Big( \langle{u}_{a,b}^\text{M} \rangle + \epsilon y_c \langle u_{a,bc}^\text{M} \rangle \Big)
	+ \epsilon \langle u_{a,bc}^\text{M} \rangle  \Big( \varphi_{abi} \delta_{jc} + \frac{\p \psi_{abci} }{\p y_j} \Big) + \dots \\
	&= L_{abij}\langle {u}_{a,b}^\text{M} \rangle 
	+ \epsilon M_{abcij} \langle{u}_{a,bc}^\text{M} \rangle+ \dots
\alend
where
\begal \label{homogenized_L_M}
	L_{abij} &= \delta_{ia} \delta_{jb} + \frac{\p \varphi_{abi}}{\p y_j} 
	\ , \\
	M_{abcij} &= y_{c} \Big( \delta_{ia} \delta_{jb} + \frac{\p \varphi_{abi}}{\p y_j} \Big) 
	+ \Big( \varphi_{abi} \delta_{jc} + \frac{\p \psi_{abci} }{\p y_j} \Big) \ .
\alend
By using the latter on the left-hand side of Eq.\,\eqref{equivalence of energy}, microscale energy becomes
\begeq \label{strain energy density2}
\int_{\Omega^P} \frac12 C^\text{m}_{ijkl} u^\text{m}_{i,j} u^\text{m}_{k,l} \d V 
=
\frac12 \int_{\Omega^P} \Big( C_{ijkl}^\text{m} L_{abij} L_{cdkl} \langle{u}_{a,b}^\text{M}\rangle \langle{u}_{c,d}^\text{M} \rangle +\\
	+ \epsilon^2 C_{ijkl}^\text{m} M_{abcij} M_{defkl} \langle{u}_{a,bc}^\text{M}\rangle \langle{u}_{d,ef}^\text{M}\rangle 
	+2 \epsilon C_{ijkl}^\text{m} L_{abij} M_{cdekl} \langle{u}_{a,b}^\text{M}\rangle \langle{u}_{c,de}^\text{M}\rangle \Big) \d V =\\
	= \frac{V}{2}\Big( \bar{C}_{abcd} \langle{u}_{a,b}^\text{M}\rangle \langle{u}_{c,d}^\text{M}\rangle 
	+ \bar{D}_{abcdef}\langle{u}_{a,bc}^\text{M}\rangle \langle{u}_{d,ef}^\text{M}\rangle 
	+ \bar{G}_{abcde}\langle{u}_{a,b}^\text{M}\rangle \langle{u}_{c,de}^\text{M}\rangle \Big)  \ .
\eqend
with
\begeq \label{relation_to_epsilon}
\bar{C}_{abcd} = \frac{1}{V} \int_{\Omega^P}  C_{ijkl}^\text{m} L_{abij} L_{cdkl} \d V \ , \\
\bar{D}_{abcdef} = \frac{\epsilon^2}{V} \int_{\Omega^P} C_{ijkl}^\text{m} M_{abcij} M_{defkl} \d V \ , \\
\bar{G}_{abcde} = \frac{2\epsilon}{V} \int_{\Omega^P}  C_{ijkl}^\text{m} L_{abij} M_{cdekl} \d V \ .
\eqend
As we have assumed centro-symmetric materials, the rank 5 tensor vanishes, $\bar{\t G}=0$. Immediately we observe by comparing with Eq.~\eqref{strain energy for homogenized material}, 
\begeq 
	C^\text{M}_{ijlm} = \bar{C}_{ijlm} \ , \\
	C^\text{M}_{ijlm} \bar{I}_{kn} + D^\text{M}_{ijklmn} = \bar{D}_{ijklmn}  \ ,
\eqend
where 
\begeq
\bar{I}_{kn} = \int_{\Omega^P} (X_k-\overset{\text{c}}X_k)(X_n-\overset{\text{c}}X_n) \d V =\epsilon^2 \int_{\Omega^P}y_k y_n \d V \ .
\eqend
Therefore, we have generated an algorithm delivering effective parameters:
\begeq \label{homogenized_tensors}
	C^\text{M}_{abcd} = \frac{1}{V} \int_{\Omega^P}  C_{ijkl}^\text{m} L_{abij} L_{cdkl} \d V \ , \\
  D^\text{M}_{abcdef} = \epsilon^2 \Bigg( \frac{1}{V} \int_{\Omega^P} C_{ijkl}^\text{m} M_{abcij} M_{defkl} \d V  - 	C^\text{M}_{abef} \int_{\Omega^P}y_c y_d \d V  \Bigg)\ ,
\eqend
after computing $\t\varphi$ and $\t\psi$ in an RVE.

\section{Numerical solution of strain gradient homogenization problems} \label{sec:3}

The overarching aim is to obtain classical stiffness tensors $C^\text{M}_{ijlm}$ and strain gradient stiffness tensors $D^\text{M}_{ijklmn}$, for their determination we need to solve Eqs.\,\eqref{diff1}, \eqref{diff2}. We restrict the analysis to 2D case, for the sake of simplicity such that all indices belong to $\{1,2\}$. Within the RVE, which is the computational domain, $\Omega$, Eqs.\,\eqref{diff1}, \eqref{diff2} are solved by using the \textsc{Galerkin} procedure in the FEM with continuous shape functions. All boundaries are given as periodic boundary conditions, in other words, the values of $\phi_{abi}$, $\psi_{abci}$ are given by \textsc{Dirichlet} boundary conditions. 

Indeed, the solutions of Eqs.\,\eqref{diff1}, \eqref{diff2} are determined for specific $a$, $b$ indices (classical coefficients) as well as $a$, $b$, $c$ indices (strain gradient coefficients). 
Consider the case that $a=1$ and $b=1$ leading to the weak form of Eqs.\,\eqref{diff1} after multiplying by an arbitrary test function vanishing on \textsc{Dirichlet} boundaries and integrating by parts
\begeq
\int_\Omega  \bigg( C_{ijkl}^\text{m} \Big(  \frac{\p \varphi_{11k}}{\p y_l} + \delta_{1k} \delta_{1l} \Big) \bigg) \frac{\partial \del \varphi_{11i}}{\partial  y_j} \d V = 0 \ ,
\eqend
out of which we determine $\varphi_{abi}$ after solving for $ab={11,12,21,22}$. By knowing $\t\varphi$, for example in the case of $a=1$, $b=1$, and $c=2$, we then solve 
\begeq
\int_\Omega \Bigg( \bigg( C_{ijkl}^\text{m} \Big(  \frac{\p \psi_{112k}}{\p y_l} +  \varphi_{11k} \delta_{l2} \Big) \bigg) \frac{\partial \del \psi_{112i}}{\partial  y_j} 
- \\
-  C_{i2kl}^\text{m} \Big(  \frac{\p \varphi_{11k}}{\p y_l} + \delta_{k1} \delta_{l1}  \Big) \del \psi_{112i}
+ {C}_{i211}^\text{M} \del \psi_{112i} \Bigg) \d V= 0 \ .
\eqend
The result in $\t\psi$ after solving for $abc=\{111, 112, 121, 122, 211, 212, 221, 222\}$. By inserting $\t\varphi$ and $\t\psi$ in Eq.\,\eqref{homogenized_L_M} and then applying Eq.\,\eqref{homogenized_tensors}, we determine $\t C^\text{M}$ and $\t	D^\text{M}$.

We have used the open-source software FEniCS for establishing the computations. The CAD models of the RVE have been created on the open source platform SALOME 7.6, and FEM discretizations of the CAD models were realized by a mesh generator NetGen built in SALOME 7.6. Applying the periodic conditions and building matrices are done via Python. We stress that the generated mesh has to possess perfectly matching vertices on opposite (periodic) boundaries for consistency. By NetGen, this has been automatically fulfilled by mapping the meshes between periodic surfaces. The mesh is transferred to the FEniCS and the numerical solution of weak forms have been obtained by using the iterative solver \emph{gmres} with preconditioner \emph{jacobi} with relative tolerance $10^{-5}$ and absolute tolerance $10^{-10}$ to ensure the accuracy of the calculations. 

\section{Identification of the classical and strain gradient stiffness tensors} \label{sec:4}

In order to demonstrate the approach, the classical and strain gradient stiffness tensors are identified for a specific cases. Firstly, it is examined for its consistency by computing $\t C^\text{M}$ and $\t	D^\text{M}$ in the case of a homogeneous material. As expected, the approach delivers zero for $\t	D^\text{M}$ (within the numerical tolerance). Concretely, the implementation delivers for $\t D^\text{M}$ components $10^{-6}$ N or smaller for a material with \textsc{Young}'s modulus of 100\,MPa and \textsc{Poisson}'s ratio of 0.3. This is consistent with the interpretation that for a homogeneous material all corresponding strain gradient material parameters vanish. 

Then a simple geometry, the so-called square lattice structure in 2D is investigated. The square lattice structure has been widely used in engineering practice \cite{arabnejad2013mechanical}, as shown in Fig.\,\ref {Yang_Homo_3} where gray lines build up a truss like structure. This inner structure is expected to deliver $D_4$ invariant material symmetry group \cite{auffray2009derivation,auffray2015complete,placidi2017identification}. 

As the microscale material parameters for the lattice structure, isotropic material properties are used with \textsc{Young}'s modulus, $E$, and \textsc{Poisson}'s ratio, $\nu$, as follows:
\begeq
C_{ijkl}^\text{m} = \lambda \delta_{ij} \delta_{kl} + \mu \delta_{ik} \delta_{jl} + \mu \delta_{il} \delta_{jk} \ , \\
\lambda = \frac{E \nu}{(1+\nu) (1-2\nu)} \ , \
\mu = \frac{E}{2(1+\nu)} \ .
\eqend
\textsc{Voigt}'s notation is used for representing the tensors, for convenience, we refer to Table~\ref{tab:Voigt_strain} and Table~\ref{tab:Voigt_strain_gradient} for the chosen convention based on the work by \cite{auffray2015complete}. 
\begin{table}
	\centering
	\caption{\textsc{Voigt}'s notation used for 2D strain tensors.}
	\begin{tabular}{p{1cm}p{1cm}p{1cm}p{1cm}p{1cm}}
		\toprule
		& $I$ & 1 & 2 & 3  \\
		\midrule
		& $ij$  & 11 & 22& 12    \\
		\bottomrule
	\end{tabular}
	\label{tab:Voigt_strain}
\end{table}
\begin{table}
	\centering
	\caption{\textsc{Voigt}'s notation used for 2D strain-gradient tensors.}
	\begin{tabular}{p{1cm}p{1cm}p{1cm}p{1cm}p{1cm}p{1cm}p{1cm}p{1cm}p{1cm}}
		\toprule
		& $I$ & 1 & 2 & 3 & 4 & 5 & 6 \\
		\midrule
		& $ijk$  & 111 & 221& 122 & 222 & 112 & 121  \\
		\bottomrule
	\end{tabular}
	\label{tab:Voigt_strain_gradient}
\end{table}

\subsection{Parameter determination for the square lattice structure} \label{lattice}

\begin{figure}
	\centering
	\includegraphics[scale=1.3]{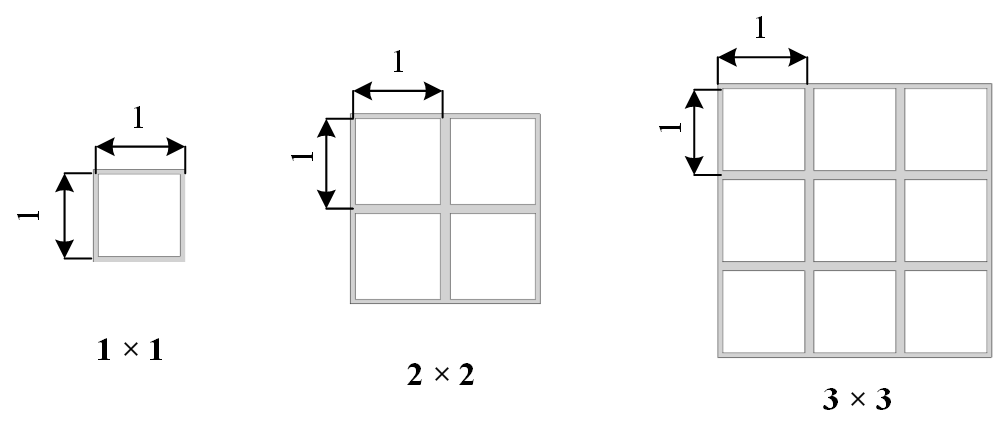}
	\caption{Geometry of square lattice structures and different selections of RVE.}
	\label{Yang_Homo_3}
\end{figure}

In the case of the square lattice structure, we consider the case that the material parameters of the inclusion are much smaller than those of the matrix, simply stated it is a additively manufactured truss like structure with rods out of a polymer and voids called inclusions. By choosing material properties as compiled in Table~\ref{tab:3} 
\begin{table}
	\centering
	\caption{Material properties used in lattice structures.}
	\begin{tabular}{lcc}
		\toprule
		Type & $E$ in MPa & $\nu$   \\
		\midrule
		Matrix  & 100.0 & 0.3    \\
		Inclusion  & $10^{-30}$ & $10^{-30}$   \\
		\bottomrule
	\end{tabular}
	\label{tab:3}
\end{table}
and the volume fraction of the inclusion as $81\%$, we select different RVEs and determine the parameters. The RVEs are generated by repeating the corresponding basic cell, and the size of basic cell is kept constant; namely the RVEs are constituted by one cell, four cells, and nine cells, as depicted in Fig.\,\ref {Yang_Homo_3}.
The results for
\begeq
\t C^\text{M} = 
\begin{pmatrix}
C_{1111} &  C_{1122}  &   0     \\
&  C_{1111}  &    0       \\
\text{sym.}       &             &  C_{1212}   \\
\end{pmatrix}
\ , \\
\t	D^\text{M} =
\begin{pmatrix}
D_{111111}& D_{111221} & D_{111122} &0 & 0 & 0    \\
&  D_{221221} &  D_{221122} & 0 & 0  & 0      \\
&             &  D_{122122} & 0 & 0  & 0 \\
&             &             & D_{111111} & D_{111221}  & D_{111122}    \\
&\text{sym.}  &             &            & D_{221221} & D_{221122}  \\
&             &             &            &             & D_{122122} \\
\end{pmatrix}
\ .
\eqend
are compiled in Table~\ref{tab:4}.
\begin{table}
	\centering
	\caption{Parameters determined for the square lattice. }
	\begin{tabular}{ccc}
		\toprule
		$C_{1111}$ in MPa 
		& $C_{1122}$ in MPa 
		& $C_{1212}$ in MPa \\
		\midrule
		11.177 
		&  0.555 
		& 0.060  \\
		\bottomrule		
		\toprule
		$D_{111111}$ in N 
		& $D_{111221}$ in N 
		& $D_{111122}$ in N \\
		\midrule
		0.005379 
		& 0.042197 
		& -0.047860 \\
		\bottomrule
		\toprule
		$ D_{221221} in N $ 
		& $D_{221122}$ in N 
		& $D_{122122}$ in N  \\
		\midrule
		1.597997 
		& 0.076341 
		& 0.033462 \\
		\bottomrule
	\end{tabular}
	\label{tab:4}
\end{table}

\begin{figure}
	\centering
	\includegraphics[scale=1.0]{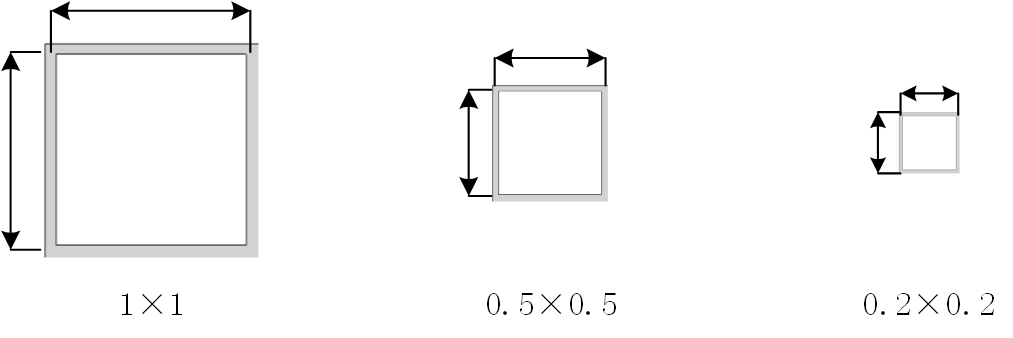}
	\caption{Different sizes of basic cell with the same volume ratio}
	\label{Yang_Homo_4}
\end{figure}

In order to investigate how the size of the basic cell affects classical and strain gradient stiffness tensors, different sizes of basic cells ($0.2\times 0.2$, $0.5\times 0.5$ ) are selected and corresponding results are compared with those obtained with the basic cells size $1\times 1$, see Fig.\,\ref {Yang_Homo_4} for the basic cells.  
Due to the fact that these three structures share the same topology, the same material properties,  as well as the same inclusion volume fraction, the corresponding classical stiffness tensors are identical. However, this is not the case for strain gradient stiffness tensors, as compiled in Table~\ref{tab:D_three square}. 
\begin{table}
	\centering
	\caption{Identified non-zero strain gradient stiffness parameters for the square lattice structure in N.}
	\begin{tabular}{ccccccc}
		\toprule
		Type & $D_{111111}$ in N & $D_{111221}$ in N & $D_{111122}$ in N & $ D_{221221}$ in N & $D_{221122}$ in N & $D_{122122}$ in N  \\
		\midrule
		1$\times$1  & 0.005379 & 0.042197 & -0.047860 & 1.597997 & 0.076341 & 0.033462  \\
		0.5$\times$0.5  & 0.001344 & 0.010549 & -0.011965 & 0.399499 & 0.019085 & 0.008365   \\
		0.2$\times$0.2  & 0.000215 & 0.001688 &  -0.001914 & 0.063919 & 0.003054 & 0.001385   \\
		\bottomrule
	\end{tabular}
	\label{tab:D_three square}
\end{table}
All non-vanishing parameters approach zero as the size of basic cells is decreasing. We remark that this fact is intuitively correct. Indeed, when the size of basic cells vanishes, the material becomes homogeneous resulting in vanishing $\t D^\text{M}$. This computation also illustrates the role of the homothetic ratio $\epsilon$. To this end, let us consider the parameter $D_{221221}$ as shown in Table~\ref{tab:D_three square}. In the case of a basic cell 1$\times$1, such a parameter is 4 times larger than that computed for the case of a basic cell 0.5$\times$0.5, and it is 25 times larger than that computed for the case of a basic cell 0.2$\times$0.2. The magnification factors (4 or 25) are equal to the square of homothetic ratios of these three basic cells as directly seen in Eq.\,\eqref{relation_to_epsilon}. 

\section{Computational validation of determined parameters} \label{sec:5}

In order to verify and validate the numerical values of the determined parameters, we perform three distinct computations: a computation at the microscale by incorporating the inner structure, a computation only with the determined classical stiffness tensor at the macroscale by using the homogenized structure and another computation with both determined classical stiffness tensor as well as strain gradient tensor at the macroscale by using the homogenized structure. 

As suggested in \cite{niiranen2016variational,rudraraju2014three,fischer2011isogeometric}, the problem of strain gradient elasticity is solved by using a weak form that, in the linear setting, leads to the $H^2$ norm about the trial solutions as well as test functions. Hence, the corresponding finite-dimensional approximations are guaranteed to lie in a function space which is at least of $C^1$ continuity. In order to obtain this property, the isogeometric FEM is employed with non-uniform rational Bezier splines (NURBS) based shape functions. The isogeometric FEM is able to ensure $C^n$ continuity in one single patch, which is appropriate for 2D simple geometries as being the case here. A detailed discussion of the NURBS basis and isogeometric FEM as well as the weak formulation of strain gradient elasticity can be found in \cite{hughes2005isogeometric,kamensky2019tigar,capobianco2018time,eugster2014director,cazzani2016isogeometric,cazzani2016constitutive,cazzani2016analytical,greco2013b}. The deformation energy which quantitatively describes the overall deformation behavior of the structures  are used to compare the results.

\begin{figure}
	\centering
	\includegraphics[scale=0.7]{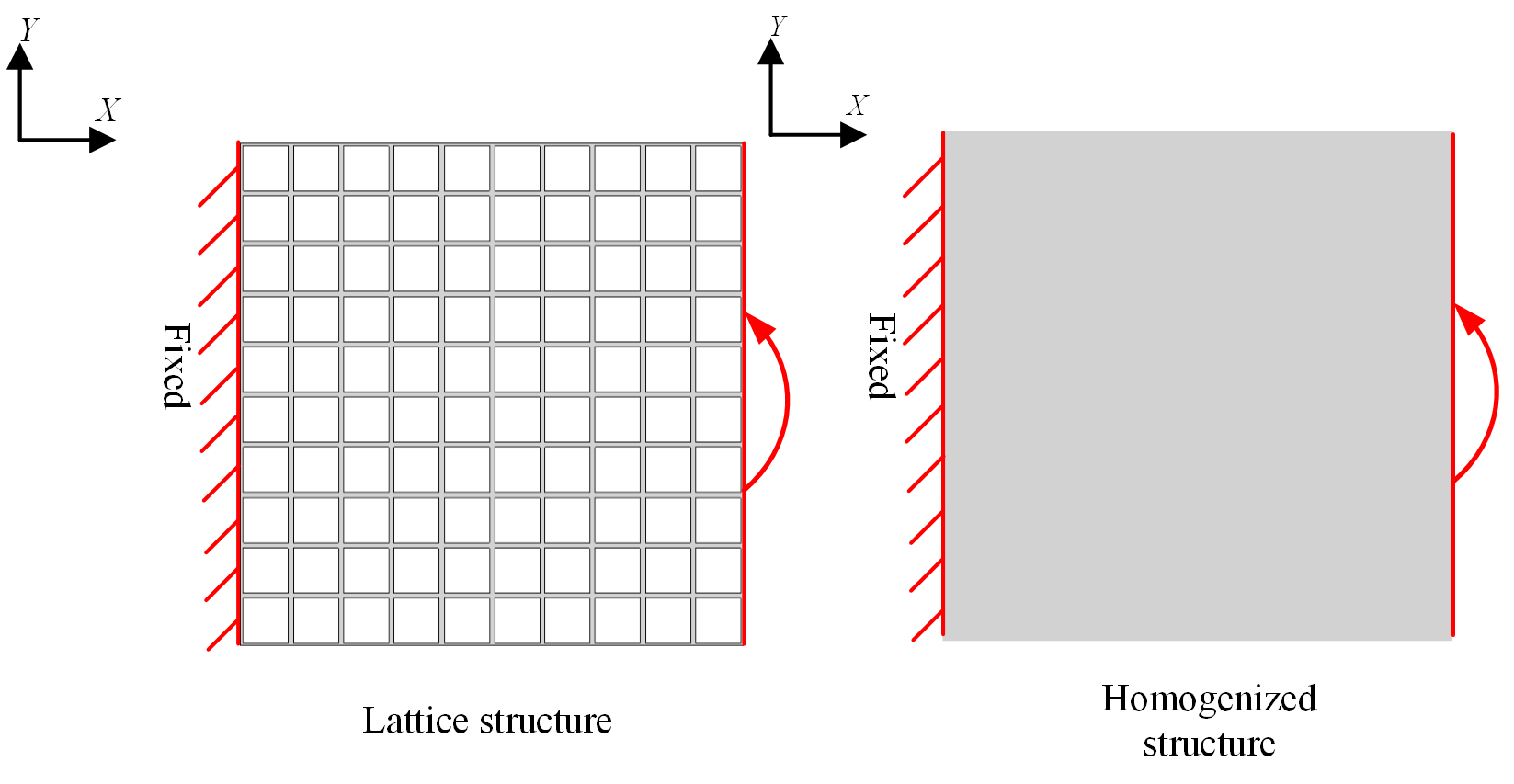}
	\caption{The boundary conditions for computations.}
	\label{Yang_Homo_5}
\end{figure}

The boundary conditions for the simulations are shown in Fig.\,\ref {Yang_Homo_5}. 
The left side of the structure is clamped and on the right side of the structure a rotation is prescribed along the center of the right edge. Two different types of computations are performed in the following subsections. In Section~\ref{5.1}, the computations are done for the lattice structures with different macro sizes but the same sizes of basic cell, and in Section~\ref{5.2} we conduct computations for lattice structures with the same macro sizes but with different sizes of internal basic cells. The total volume remains the same in this case due to the fact that the ratio of the cell wall length to thickness of the basic cell is held constant with a ratio 1 to 10.

\subsection{Computations for square lattices with the same basic cell sizes and varied macro sizes} \label{5.1}

\begin{figure}
	\centering
	\includegraphics[scale=0.6]{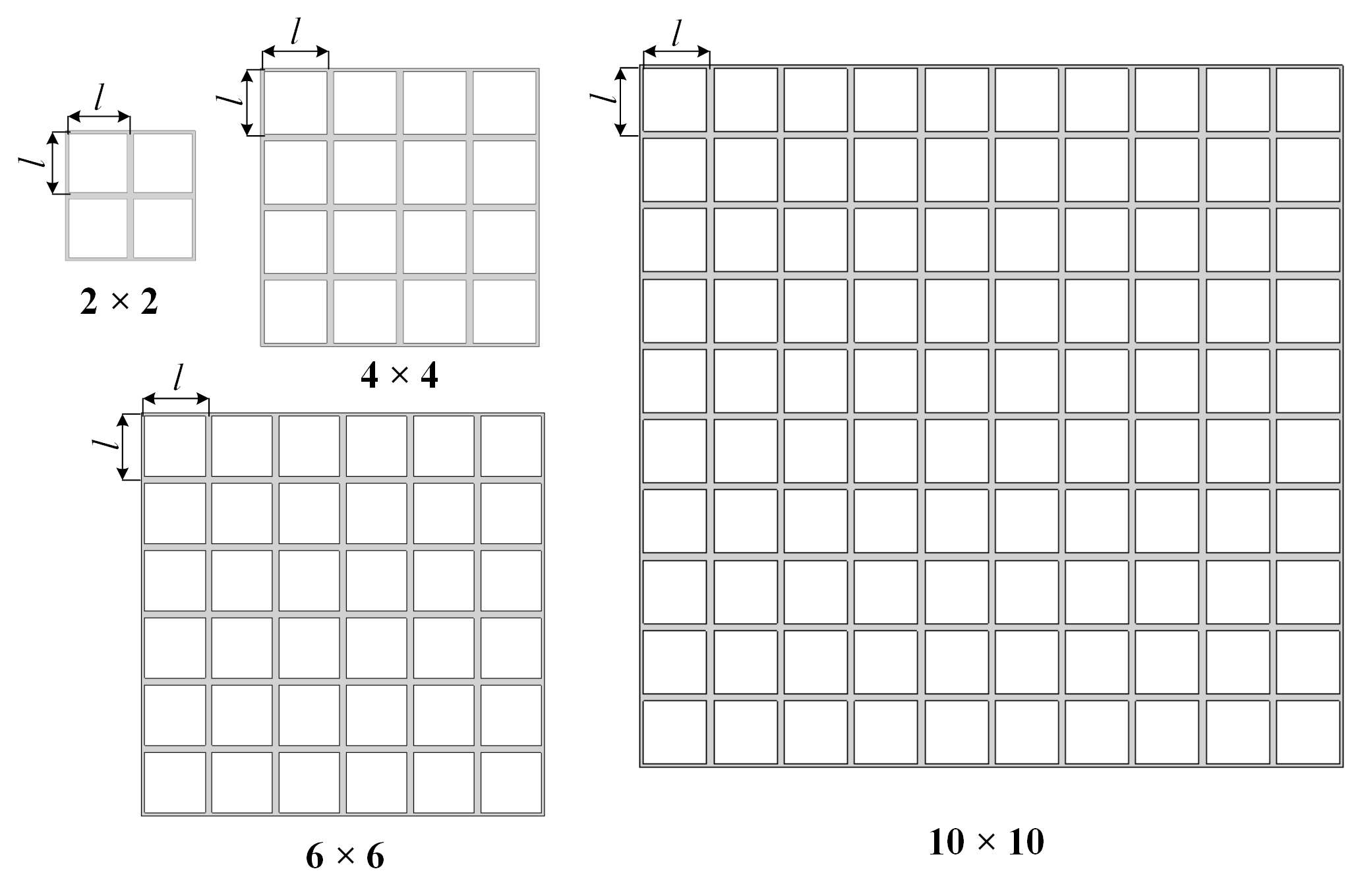}
	\caption{Selected simulations for square lattice with the same size micro-structure.}
	\label{Yang_Homo_6}
\end{figure}

In this section, the computations for the square lattice with the same sizes of the basic cell but with different macro sizes as shown in Fig.\,\ref {Yang_Homo_6} are done. The size of the basic cell is 1\,mm\,$\times$\,1\,mm, and the selected lattices are of the macro sizes
\begin{itemize}
	\item
	2\,mm\,$\times$\,2\,mm,
	\item
	4\,mm\,$\times$\,4 mm,
	\item
	6\,mm\,$\times$\,6 mm,
	\item
	10\,mm\,$\times$\,10\,mm.
\end{itemize}
 
\begin{figure}
	\centering
	\subfigure[2 $\times$ 2 ]{\includegraphics[width=0.475\textwidth]{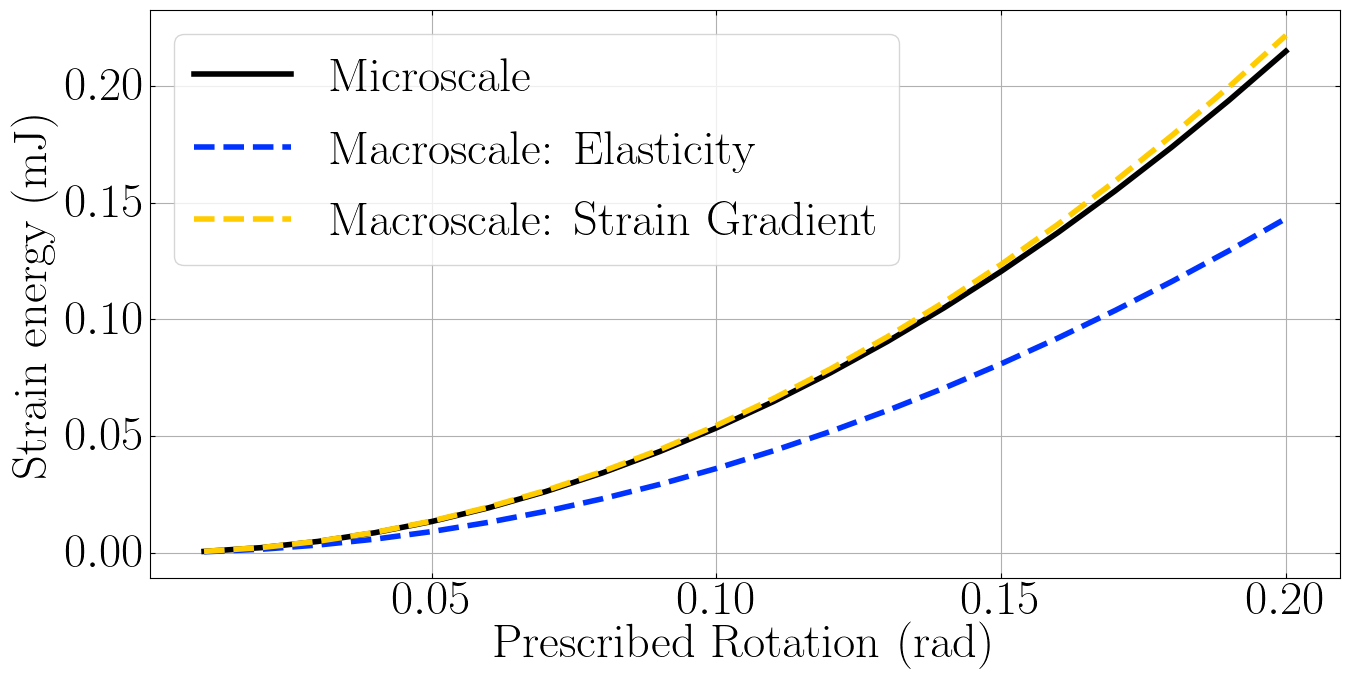}}
	\subfigure[4 $\times$ 4]{\includegraphics[width=0.475\textwidth]{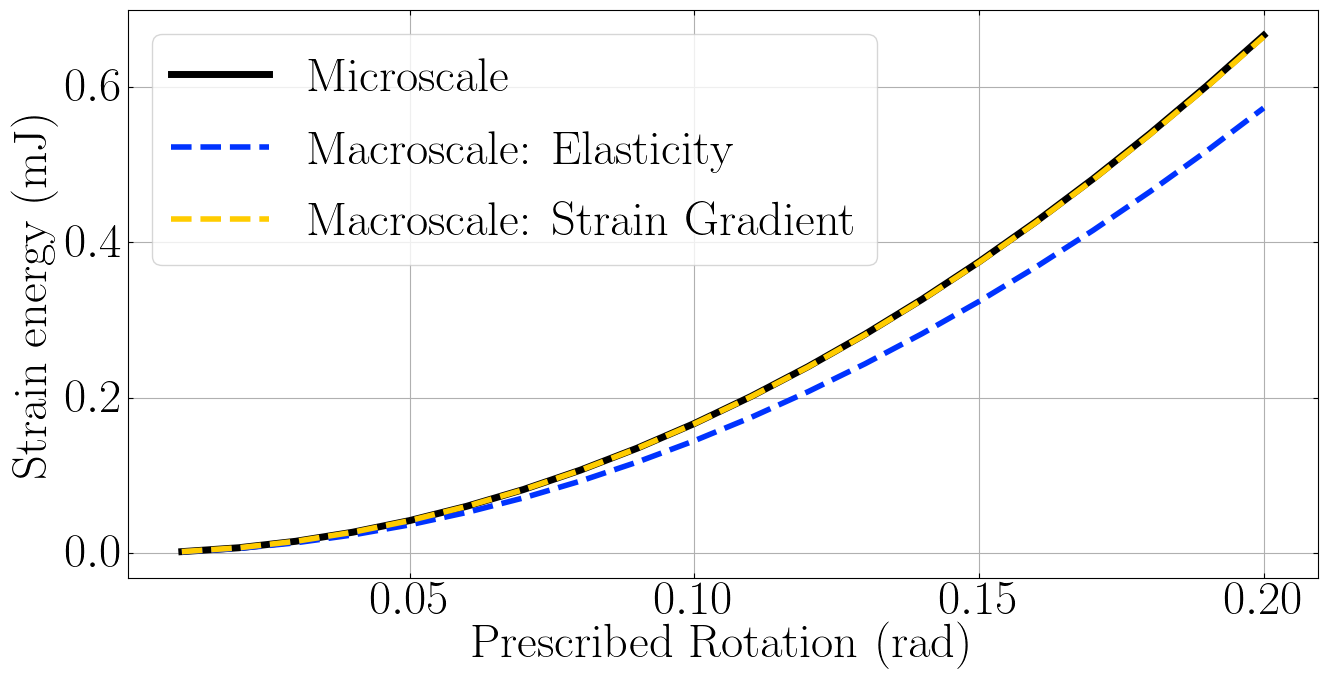}}
	\subfigure[6 $\times$ 6]{\includegraphics[width=0.475\textwidth]{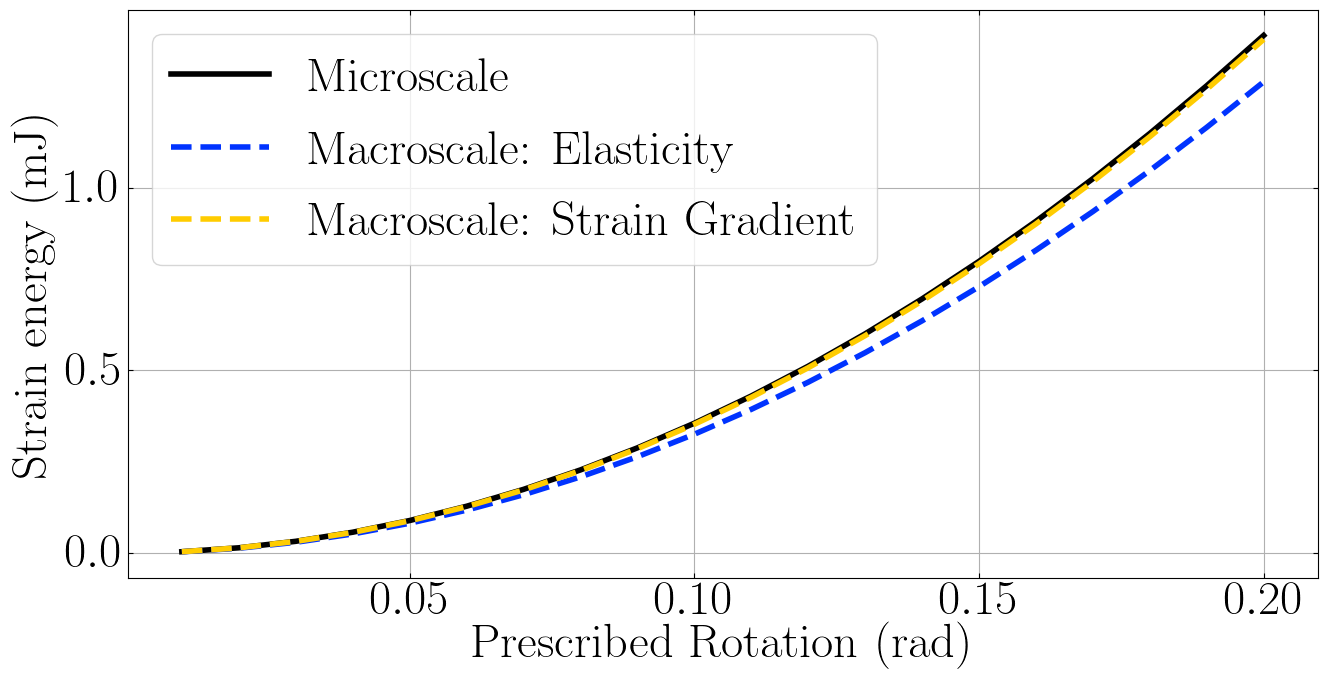}}
	\subfigure[10 $\times$ 10]{\includegraphics[width=0.475\textwidth]{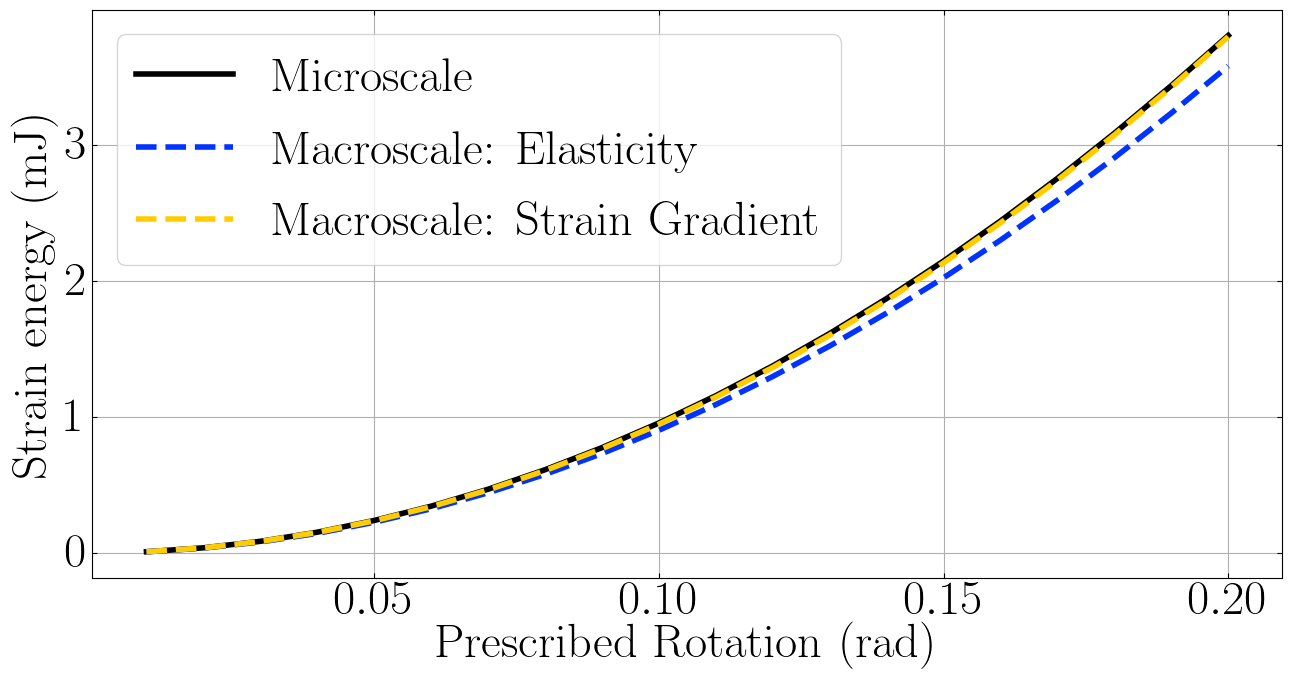}}
	\caption{ Comparisons of strain energies for square lattice structures with different macroscale sizes.} 
	\label{Yang_Homo_7}
\end{figure}

The results of the simulations are shown in Fig.\,\ref {Yang_Homo_7}, where the vertical axis stands for the strain energy of the structures (in mJ) and the horizontal axis stands for the prescribed rotation (in rad).  The black solid lines in Fig.\,\ref {Yang_Homo_7} represents the results from the microscale. We understand this solution as the correct one. The blue dashed line represents the computations of the homogenized structure by using the classical stiffness tensor. The yellow dashed line represents the simulations for the homogenized structure taking the strain gradient effect into account. 

The blue lines show a smaller strain energy with regard to the microscale due to the absence of the higher order strain gradient energy. We remark that, keeping the sizes of the basic cells unchanged, with increasing macro sizes of the structures, namely $L/l$ becoming larger and larger, the computations under classical elasticity theory approach to that of microscale. We may say that in a large macro scale $L/l>10$, the classical elasticity is adequate to guarantee the accuracy of the computation. However, when the macroscopic length scale is of the same order of its sizes of internal substructures, the strain gradient effect becomes significant. This phenomenon is also known as \textit{size effect}.

\subsection{Computations for square lattices with varied basic cell sizes and the same macro sizes}\label{5.2}

\begin{figure}
	\centering
	\includegraphics[scale=0.6]{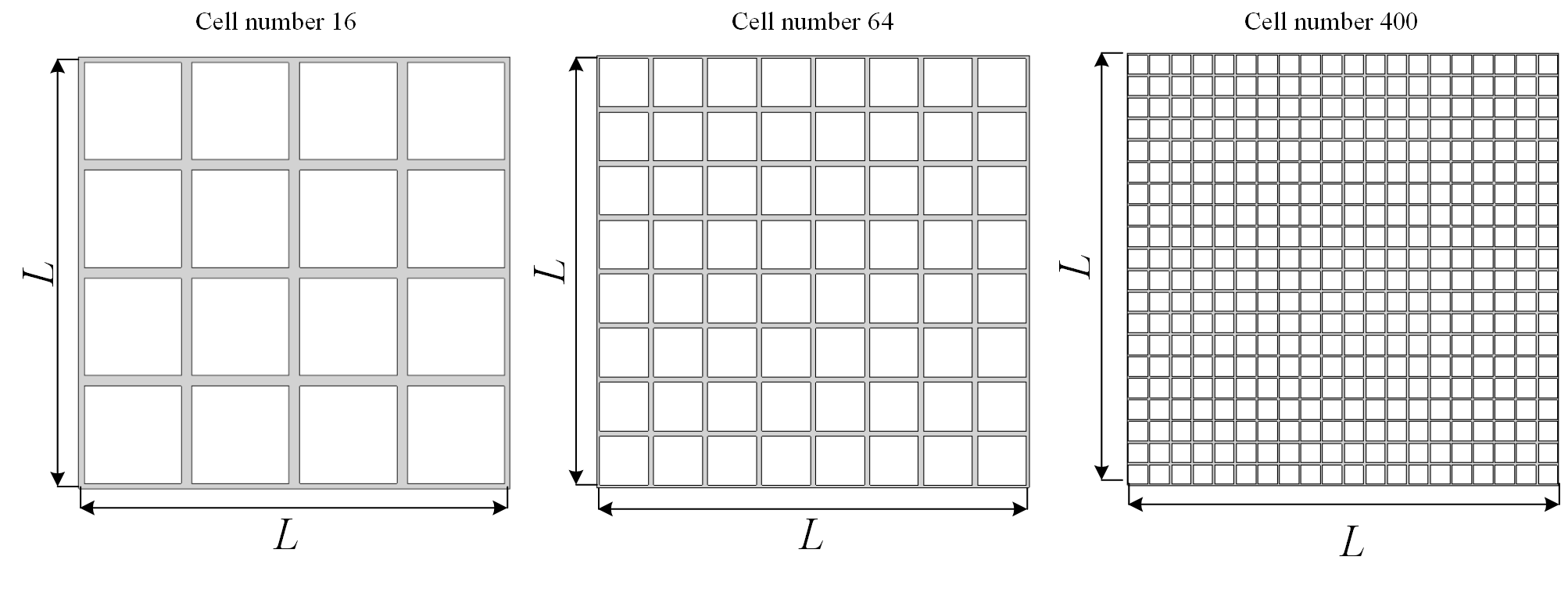}
	\caption{Selected simulations for square lattices with basic cells of varied sizes.}
	\label{Yang_Homo_8}
\end{figure}

\begin{figure}
	\centering
	\includegraphics[scale=0.25]{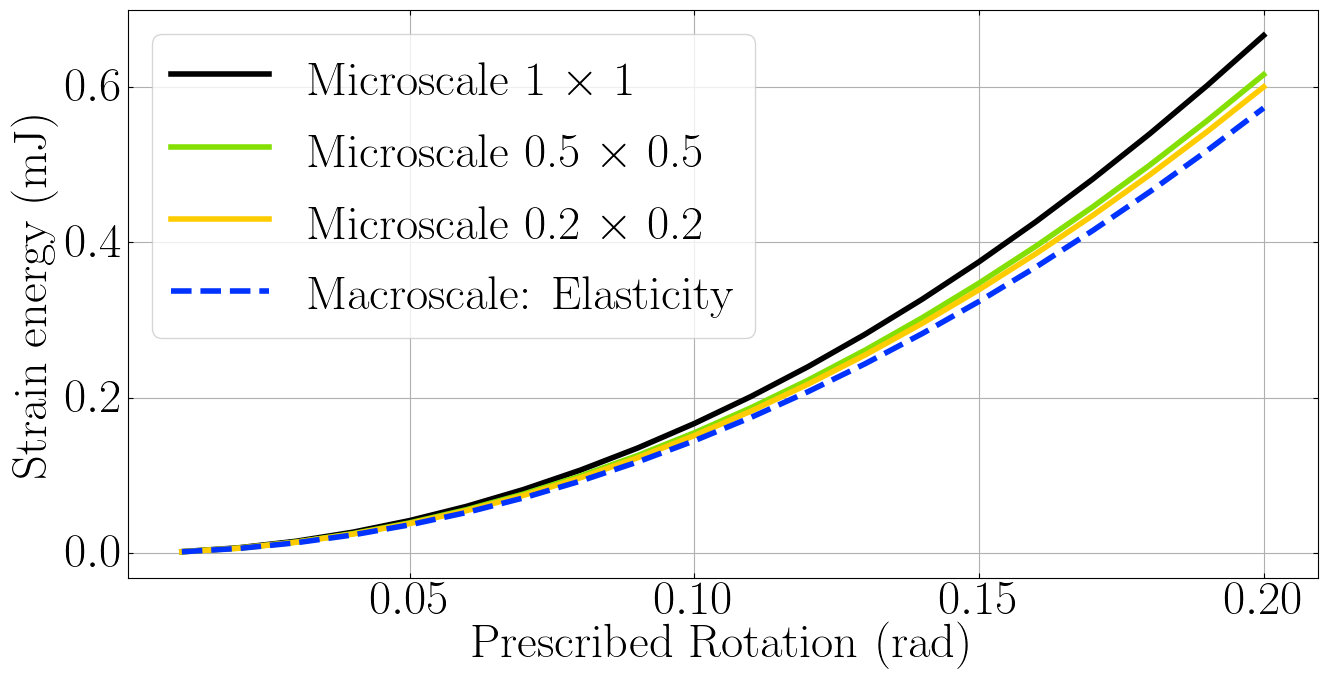}
	\caption{Comparative computations between microscale and macroscale of elasticity.}
	\label{Yang_Homo_9}
\end{figure}

\begin{figure}
	\centering
	\subfigure[Comparisons for structures with 16 cells ]{\includegraphics[width=0.475\textwidth]{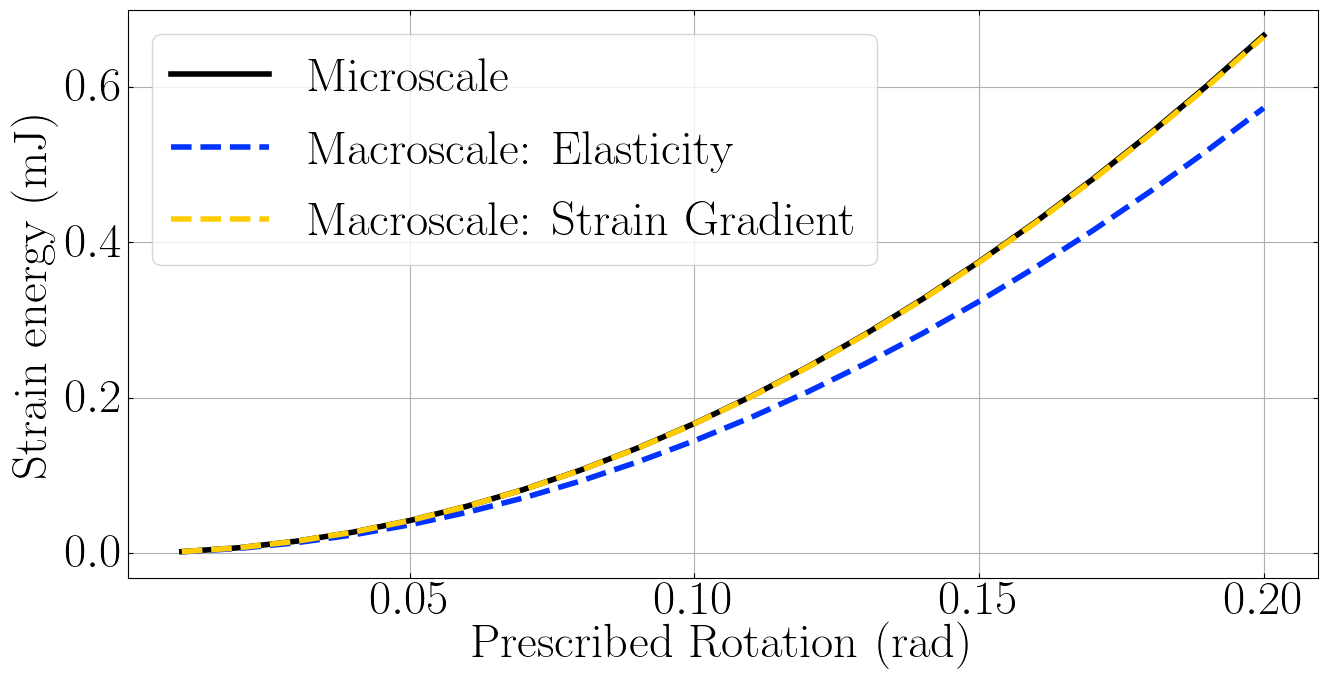}}
	\subfigure[Comparisons for structures with 64 cells]{\includegraphics[width=0.475\textwidth]{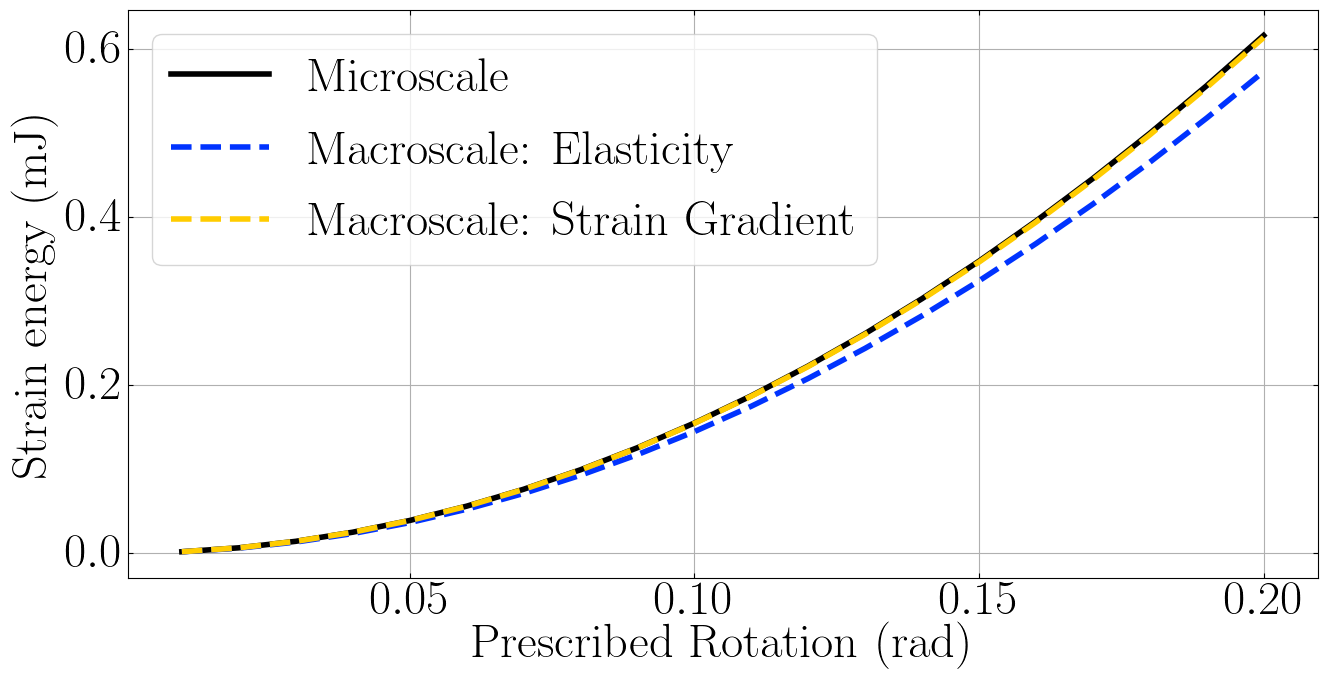}}
	\subfigure[Comparisions for structures with 400 cells]{\includegraphics[width=0.475\textwidth]{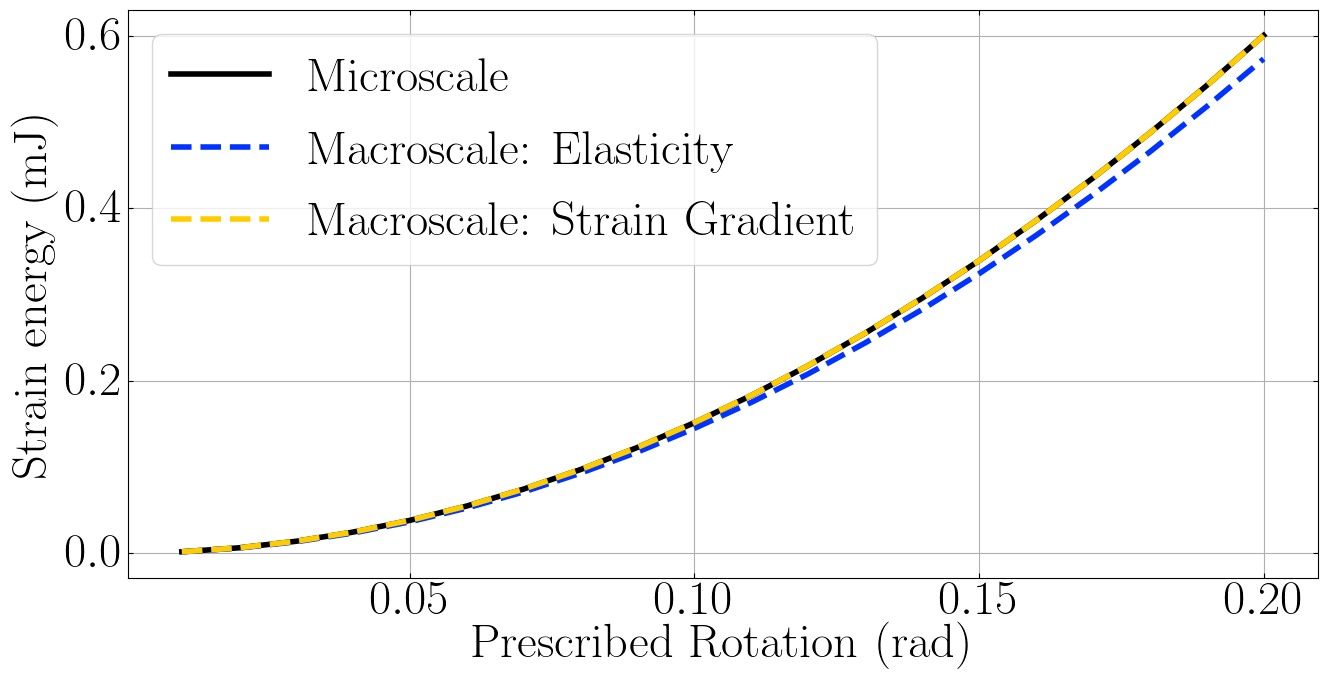}}
	\caption{ Comparisons between computations in microscale, macroscale of elascity, and macroscale of strain gradient for square lattice structures with different number of cells.} 
	\label{Yang_Homo_10}
\end{figure}

In this section, in order to further verify the identified parameters for square lattices with different basic cell sizes but the same macro sizes, computations are conducted herein. Three square lattices are selected as shown in Fig.\,\ref {Yang_Homo_8}. These three lattices possess the same macro sizes 4 mm $\times$ 4 mm,  and the basic cell sizes of them are 1 mm $\times$ 1 mm, 0.5 mm $\times$ 0.5 mm and 0.2 mm $\times$ 0.2 mm for the left, the middle and the right lattice in Fig.\,\ref {Yang_Homo_8}, respectively, which divides the macro domain into 16, 64 and 400 basic cells. 
The computations are shown in Fig.\,\ref {Yang_Homo_9} and Fig.\,\ref {Yang_Homo_10}. 
Fig.\,\ref {Yang_Homo_9} indicates that with the increasing of  basic cell sizes, the strain energy at the rotation of 0.2 radian shows an increasing trend shown under microscale. This scale-dependent (depends on $L/l$) phenomenon is also know as size effect. 
The computations for macroscale of elasticity are identical for these three cases due to the fact that ratio of the cell wall length to thickness of the basic cell is fixed with a ratio 1 to 10.  The computation for macroscale of elasticity are independent of the scale ratio and they show a significant error compared with the microscale where the scale ratio ($L/l$) is getting smaller, and the size effect could not be ignored. When the scale ratio ($L/l$) is getting larger, which means the decreasing of the basic cell sizes, the computation for macroscale of elasticity are gradually approaching to the microscale. In such a case, for example $L/l>20 $, the size effect can be ignored. It can be also observed from Fig.\,\ref {Yang_Homo_10} (a), (b), (c) that the computations with strain gradient show a good match with the microscale quantitatively, which means by taking the strain gradient stiffness tensor into account, the size effect of the lattice structure is fully resolved.

\section{Conclusions}
\label{sec:conclusions}
A homogenization approach based on the asymptotic analysis has been exploited for developing a methodology to determine parameters in a metamaterial. Specifically, the strain gradient theory is used at the macroscale. The expressions of classical stiffness tensor and strain gradient stiffness tensor have been derived and the FEM has been successfully used to solve the partial differential equations generated from the homogenization procedure. The so-called square lattice structure has been investigated, and their material parameters are explicitly computed. The proposed approach guarantees that the parameters of strain gradient stiffness tensors vanish as the material becomes homogeneous. Moreover, it ensures that strain gradient related parameters are independent on the repetition of RVE, but dependent on the intrinsic size of the material. In order to validate the parameters determined by this methodology, additional numerical computations of the square lattice with different sizes have been performed. The numerical results show that the size effect of the lattice can be accurately captured by using the strain gradient theory with the parameters determined by the methodology applied herein. We stress that this methodology simply allows for any metamaterial made of a substructure with an RVE.

\section*{Appendix: Asymptotic solution for the displacement field}
The asymptotic solution for an RVE are derived. Namely the solutions of Eq.\,\eqref{first terms}, Eq.\,\eqref{second terms} and Eq.\,\eqref{third terms} are shown.

We start with Eq.\,\eqref{first terms}. As $C_{ijkl}^\text{m}$ is a function of $\t y$, the only possible general solution of Eq.\,\eqref{first terms} is to restrict $\overset{0}u_i(\t X)$ since it is $\t y$-periodic and has a bounded gradient. The solution in the order of $\epsilon^{-2}$ can be given as

\begin{equation} \label{first terms solution}
	\overset{0}u_i = \overset{0}u_i(\t X) \ .
\end{equation}

Note that $\overset{0}u_i(\t X)$ depends only on the macroscopic coordinates; it is assumed to be the known macroscopic displacement $\overset{0}{u}_i(\t X) =u^\text{M}_i(\t X)$. By substituting Eq.\,\eqref{first terms solution} into Eq.\,\eqref{second terms}, by introducing $\varphi_{abc}=\varphi_{abc}(\t y)$, for the inverse operation, we obtain
\begeq \label{diff.eq.phi}
\frac{\partial C_{ijab}^\text{m}}{\partial y_j} \frac{\partial \overset{0}u_a}{\partial X_b} 
= - \frac{\partial }{\partial  y_j} \bigg( C_{ijkl}^\text{m}   \frac{\partial \overset{1}u_k}{\partial y_l} \bigg) 
\ , \\
\frac{\partial C_{ijab}^\text{m} }{\partial  y_j}  
= - \frac{\partial }{\partial  y_j} \bigg( C_{ijkl}^\text{m} \frac{\partial \varphi_{abk} }{\partial  y_l} \bigg)
\ , \\
\frac{\partial }{\partial  y_j} \bigg( C_{ijkl}^\text{m} \Big(  \frac{\p \varphi_{abk}}{\p y_l} + \delta_{ak} \delta_{bl} \Big) \bigg)
= 0 \ .
\eqend
Then the general solution of Eq.\,\eqref{second terms} can be given as

\begin{equation} \label{second terms solution}
\overset{1}{u}_i = \varphi_{abi} \overset{0}{u}_{a,b} + \overset{1}{\bar u}_i(\t X) \ ;
\end{equation}
where $\overset{1}{\bar u}_i =\overset{1}{\bar u}_i(\t X)$ are integration constants in $\t y$.

Substituting Eq.\,\eqref{first terms solution} and Eq.\,\eqref{second terms solution} (with $\overset{1}{\bar u}_i(\t X) = 0$) into Eq.\,\eqref{third terms} leads to 

\begeq \label{third terms derivation}
C_{ijkl}^\text{m}   \overset{0}u_{k,lj}  
+  C_{ijkl}^\text{m}   \frac{\partial \varphi_{abk}}{\partial y_l} \overset{0}u_{a,bj}
+  \frac{\partial }{\partial  y_j} \big( C_{ijkl}^\text{m} \varphi_{abk}  \big) \overset{0}u_{a,bl}
+ \frac{\partial }{\partial  y_j} \Big( C_{ijkl}^\text{m}  \pd{\overset{2}u_k}{y_l} \Big)   
+ f_i = 0 \ .
\eqend

Please note that the body force $\t f$ keeps unchanged in micro- and macro-scales . We recall the governing equation in the macro-scale which reads \cite{021}
\begeq \label{governing equation}
\Bigg(\frac{\p w^\text{M}}{\p u_{i,j}^\text{M}} - \Big(\frac{\p w^\text{M}}{\p u_{i,jk}^\text{M}}\Big)_{,k} \Bigg)_{,j}  + f_i= 0 \ , \\
C^\text{M}_{ijkl}  u^\text{M}_{k,lj}
- D_{ijklmn}^\text{M}    u_{l,mnkj}^\text{M}   + f_i= 0 \ .
\eqend
By neglecting the fourth order term in Eq.\,\eqref{governing equation} and using $\overset{0}{u}_i(\t X) =u^\text{M}_i(\t X)$, we obtain
\begeq \label{body force}
f_i =  -C^\text{M}_{ijkl}  u^\text{M}_{k,lj} =  -C^\text{M}_{ijkl}  \overset{0}u_{k,lj} \ .
\eqend

Substituting Eq.\,\eqref{body force} into Eq.\,\eqref{third terms derivation} leads to

\begeq \label{third terms derive}
\frac{\partial }{\partial  y_j} \Big( C_{ijkl}^\text{m}  \pd{\overset{2}u_k}{y_l} \Big)   
= -\Big( C_{icab}^\text{m} + C_{ijkl}^\text{m}   \frac{\partial \varphi_{abk}}{\partial y_l} \delta_{jc}  +  \frac{\partial }{\partial  y_j} \big( C_{ijkl}^\text{m} \varphi_{abk}  \big) \delta_{lc} - C^\text{M}_{icab} \Big) \overset{0}u_{a,bc} \ . \\ 
\eqend
As $\overset{0}u_{a,bc}$ is constant in $\t y$, we can introduce $\psi_{abci}$ depending on $\t y$ and decompose as follows:

\begeq \label{third terms solution}
\overset{2}{u}_i = \psi_{abci} \overset{0}{u}_{a,bc} + \overset{2}{\bar u}_i(\t X) \ .
\eqend
where $\psi_{abcd}=\psi_{abcd}(\t y)$ and $\overset{2}{\bar u}_i(\t X)$ are integration constants in $\t y$. By substituting Eq.\,\eqref{third terms solution} (with $\overset{2}{\bar u}_i(\t X) = 0$) into Eq.\,\eqref{third terms derive}, it is found that the tensor $\psi_{abcd}$ must fulfill the following equation
\begeq  \label{diff.eq.psi}
\frac{\partial }{\partial  y_j} \bigg( C_{ijkl}^\text{m} \Big(  \frac{\p \psi_{abck}}{\p y_l} +  \varphi_{abk} \delta_{lc} \big) \bigg)
+  C_{ickl}^\text{m} \Big(  \frac{\p \varphi_{abk}}{\p y_l} + \delta_{ka} \delta_{lb}  \Big) - {{C}}_{icab}^\text{M} = 0  \ ,
\eqend
such that Eq.\,\eqref{displacement function} provides
\begin{equation} 
u^\text{m}_i(\t X,\t y) 
= \overset{0}{u}_i(\t X) 
+ \epsilon \varphi_{abi}(\t y) \overset{0}{u}_{a,b}(\t X) 
+ \epsilon^2 \psi_{abci}(\t y) \overset{0}{u}_{a,bc}(\t X) 
+ \dots \ .
\end{equation}

\section*{Acknowledgements}
We express our gratitude to Emilio Barchiesi, Ivan Giorgio, and Francesco dell'Isola for valuable discussions. We also thank David Kamensky for the help of implementation of isogeometric FEM in FEniCS.

\bibliographystyle{spmpsci}
\bibliography{Yang2019_homogenization}

\end{document}